# Identification of Repurposable Drugs and Adverse Drug Reactions for Various Courses of COVID-19 Based on Single-Cell RNA Sequencing Data

Zhihan Wang[1]*, Kai Guo[2]*, Pan Gao[2,3]*, Qinqin Pu[2], Min Wu[2$], Changlong Li[1$], and Junguk Hur[2$]

[1] West China School of Basic Medical Sciences & Forensic Medicine, Sichuan University, Chengdu, Sichuan 610041, China

[2] Department of Biomedical Sciences, University of North Dakota School of Medicine and Health Sciences, Grand Forks, ND 58202, USA

[3] State Key Laboratory of Biotherapy and Cancer Center, West China Hospital, Sichuan University, and Collaborative Innovation Center for Biotherapy, Chengdu, Sichuan 610041, China

\* These authors contributed equally.

$ Correspondence:

- Dr. Min Wu, Email: min.wu@und.edu, Department of Biomedical Sciences, University of North Dakota School of Medicine and Health Sciences, Grand Forks, ND 58202, USA;
- Dr. Changlong Li, Email: changlongli@scu.edu.cn, West China School of Basic Medical Sciences & Forensic Medicine, Sichuan University, Chengdu, Sichuan 610041, China;
- Dr. Junguk Hur, Email: junguk.hur@med.und.edu, Department of Biomedical Sciences, University of North Dakota School of Medicine and Health Sciences, Grand Forks, ND 58202, USA.

**ABSTRACT**

Coronavirus disease 2019 (COVID-19) has impacted almost every part of human life worldwide, posing a massive threat to human health. There is no specific drug for COVID-19, highlighting the urgent need for the development of effective therapeutics. To identify potentially repurposable drugs, we employed a systematic approach to mine candidates from U.S. FDA-approved drugs and preclinical small-molecule compounds by integrating the gene expression perturbation data for chemicals from the Library of Integrated Network-Based Cellular Signatures project with a publicly available single-cell RNA sequencing dataset from mild and severe COVID-19 patients. We identified 281 FDA-approved drugs that have the potential to be effective against SARS-CoV-2 infection, 16 of which are currently undergoing clinical trials to evaluate their efficacy against COVID-19. We experimentally tested the inhibitory effects of tyrphostin-AG-1478 and brefeldin-a on the replication of the single-stranded ribonucleic acid (ssRNA) virus influenza A virus. In conclusion, we have identified a list of repurposable anti-SARS-CoV-2 drugs using a systems biology approach.

**INTRODUCTION**

Coronavirus disease 2019 (COVID-19) is a highly contagious respiratory disease resulting from a life-threatening novel coronavirus, severe acute respiratory syndrome coronavirus 2 (SARS-CoV-2). It has spread rapidly across the globe, resulting in over 42.5 million confirmed cases and 1.1 million deaths as of October 25, 2020[1, 2]. SARS-CoV-2 is an enveloped RNA virus that belongs to the genus *Betacoronavirus* of the family *Coronaviridae*, which includes well-known severe acute respiratory syndrome coronavirus (SARS-CoV) and Middle East respiratory syndrome coronavirus (MERS-CoV)[3]. Advancement in the management of these coronaviruses and other viruses, such as influenza virus H1N1 and Ebola infections, has provided insight into treating COVID-19.

More than 3,600 active clinical trials for COVID-19 are being performed[4, 5]. According to the World Health Organization (WHO), there are more than 154 candidate vaccines in preclinical evaluation and 44 candidate vaccines in phase 1 trials or beyond (last updated: October 25, 2020), several of which have entered phase 3 large-scale randomized clinical trials and shown benefits[6]. Unexpected side effects and immune-escape mutations pose a great challenge for the safety and efficacy of vaccines[7]. Chloroquine[8, 9] and its hydroxyl analog hydroxychloroquine[10], lopinavir/ritonavir[11-13], and remdesivir[9, 14], developed for treating malaria, human immunodeficiency virus (HIV), and Ebola virus, respectively, have been previously suggested to treat COVID-19 and are being tested in clinical trials. However, at present, some clinical trials have shown that these drugs have little or no effect on the treatment of hospitalized COVID-19 patients[15-19]. Hence, there is an urgent need to search for new repurposed drugs.

COVID-19 is not the first outbreak of zoonotic coronaviruses. The severe acute respiratory syndrome (SARS) outbreak, first identified in the Guangdong Province of southern China in 2002, lasted for eight months, resulting in 8,098 confirmed human cases in 29 countries, with 774 deaths (case fatality rate of 9.6%)[20, 21]. Approximately ten years later, in 2012, Saudi Arabia isolated another highly pathogenic coronavirus, MERS-CoV, from the sputum of a male patient who died of acute pneumonia and renal failure[22]. MERS-CoV caused an outbreak with 2,260 cases and 803 deaths (case fatality rate of 35.5%)[23, 24].

Although COVID-19 is less fatal than SARS and MERS, older patients with comorbidities tend to experience more severe symptoms, making them more vulnerable. Most SARS-CoV-2-infected patients display mild symptoms and generally have a good prognosis, classified as mild COVID-19[11, 25]. However, a large proportion of patients, especially older men with underlying chronic diseases, have rapidly progressed to severe COVID-19 and suffered from respiratory distress requiring emergent medical interventions[26]. Unfortunately, there are no FDA-approved vaccines or specific and effective drugs for COVID-19 yet[4, 5].

Additionally, recent studies have shown the critical roles of host immune responses in protection and the pathogenesis of respiratory viral infections, for instance, SARS-CoV, MERS-CoV, and influenza A viruses[27, 28]. Liao et al.[29] reported that an increase in $CD8^+$ T cells in COVID-19 patients correlates with an improved outcome. They also proposed therapeutic strategies by targeting the myeloid cell compartment to treat COVID-19-associated inflammation.

It is critical to find potentially useful drugs for COVID-19 among currently available drugs. Drug repurposing is an essential and universal strategy in developing new drugs and a potentially important strategy for discovering existing medicines to tackle COVID-19[30]. It may facilitate the discovery of new mechanisms of action for existing drugs, which is less time consuming and is cost effective, and moreover, the pharmaceutical supply chains for formulation and distribution already exist[31, 32]. Gordon et al.[33] also reported 69 repurposable known compounds targeting 66 human proteins based on 332 high-confidence SARS-CoV-2-human protein-protein interactions, which were physically associated by using affinity-purification mass spectrometry. Another study[34] presented data on the antiviral activity of 20 FDA-approved drugs against SARS-CoV-2 that have previously been shown to inhibit SARS-CoV and MERS-CoV. Another research team[35] conducted a high-throughput analysis of the ReFRAME library to identify 30 candidate existing drugs that prevent SARS-CoV-2 from replicating in mammalian cells. In a study[36] based on public data of patients with pulmonary fibrosis and the Library of Integrated Network-Based Cellular Signatures (LINCS)[37], several drugs targeting ACE2 were identified to be repurposable against COVID-19.

Considering that an RNA virus exhibits a considerable degree of sequence variation, drugs targeting host factors may cause less mutational resistance with more significant and broad antiviral spectrum potential[38]. Hence, there is an urgent need to identify potential therapeutics with new strategies for emerging infectious diseases. Repurposing clinically assessed drugs represents one of the most practical strategies for rapidly identifying treatments to combat COVID-19.

In this study, we analyzed a publicly available single-cell RNA sequencing (scRNA-seq) dataset of bronchoalveolar lavage fluid (BALF) collected from mild and severe COVID-19 patients as well as separate bulk RNA-seq data of BALF from COVID-19 patients (Figure 1). Data mining was performed by using the drug perturbation database LINCS to identify potential therapies for COVID-19. A total of 281 candidates with different courses of COVID-19 independent of cell subtypes were identified. Additionally, we identified and validated two candidates without adverse drug reactions (ADRs), tyrphostin-AG-1478 and brefeldin-a, exhibiting antiviral activity against the single-stranded ribonucleic acid (ssRNA) virus influenza A virus (IAV). Our findings may aid in the rapid preclinical and clinical evaluation of these therapeutics and provide an important drug discovery pipeline to accelerate and facilitate the development of potential treatments for COVID-19.

**MATERIALS AND METHODS**

*ScRNA-seq data analysis and sample aggregation*

The gene-barcode matrix files of all 6 COVID-19 donors containing 3 mild cases and 3 severe cases (lung BALF) and 3 healthy controls (lung tissues) were downloaded from the NCBI Gene Expression Omnibus database (accession ID: GSE145926)[29]. The expression matrices were loaded into the R statistical analysis platform using Seurat v3[39], keeping only the cells with gene numbers between 200 and 6,000, unique molecular identifier (UMI) count above 1,000, and mitochondrial gene percentage below 0.1. A total of 43,914 cells from 9 samples were used for our further analyses. In addition to these scRNA-seq data, we also collected a list of differentially expressed genes (DEGs) in a separate cohort of SARS-CoV-2-infected lung BALF using a bulk RNA-seq analysis to compare against the single cell-based data. This DEG list was obtained from the Chinese National Genomics Data Center (https://bigd.big.ac.cn/; accession ID: CRA002390)[40].

*Dimensionality reduction and clustering*

The LogNormalize method in Seurat was used for normalizing the filtered gene-barcode matrix. Principal component analysis (PCA) was performed by using the top 2,000 most variable genes. Uniform Manifold Approximation and Projection (UMAP) was performed on the top 50 principal components for visualizing the cells. Graph-based clustering was performed on the PCA-reduced data with Seurat.

*Differential analysis for clusters among the three groups*

MAST, a generalized linear model framework treating the cellular detection rate as a covariate, was used to perform differential analysis. DEGs were identified by comparing each cluster between all three groups. Genes with average |log2FC| > 0.25 and adjusted p-value < 0.05 were deemed DEGs.

*Drug repurposing using the LINCS drug-perturbation data*

DEGs were first sorted by the log2FC values. The upregulated and downregulated genes were then chosen to identify drugs and compounds against the LINCS database using the Connectivity Map Linked User Environment (CLUE: https://clue.io) platform[37]. The drug connectivity score (CS) with a negative value smaller than -90 was used to determine candidate drugs and compounds. The COVID-19 database from the International Clinical Trials Registry Platform (ICTRP) (https://www.who.int/ictrp/en/, updated on Oct 25$^{th}$, 2020) was searched for clinical trial information associated with these drugs.

*Adverse drug reaction analysis*

Both on-label and off-label adverse drug reactions (ADRs) of all candidate drugs with CS values smaller than -90 were collected from the SEP-L1000 database (https://maayanlab.net/SEP-L1000/). The SEP-L1000 data include on-label ADRs of FDA-approved drugs collected from SIDER[41] and off-label ADRs from the

PharmGKB database[42] based on the postmarketing ADR reports in the FDA Adverse Event Report System (FAERS).

## Cells and virus

Calu-3 and Vero E6 cells were maintained in Eagle's minimum essential medium (EMEM, Quality Biological Inc) supplemented with 10% heated-inactivated fetal bovine serum (FBS, Gibco) and penicillin and streptomycin (100 U/ml; Gibco) at 37°C with 5% $CO_2$. Influenza A virus (IAV, Puerto Rico/8/1934(H1N1)) viral stocks were titered by $TCID_{50,}$ as described previously[43]. These experiments were performed under biosafety level 2 (BSL-2) conditions.

## Antiviral assay

The drug-induced cytotoxicity of the test drugs (tyrphostin-AG-1478 and brefeldin-a) on Calu-3 and Vero E6 cells was determined by a cell counting kit 8 (CCK-8, APExBIO). Cells were cultured overnight in 96-well cell-culture Petri dishes at a density of $1 \times 10^4$ cells/well. A 10 mM stock of drug was serially diluted in 100% DMSO to obtain a 10-point dilution series. Plates were cultured with fresh drug-containing medium at 37°C for 24 h. The DMSO concentration remained constant under all conditions at 0.05%. The absorbance of each well was determined by using a 96-well multiscanner. After subtracting background absorption, the results were expressed as a percentage of viability relative to that of control cultures that received no drug. Drug concentrations that inhibited cell viability by 50% ($IC_{50}$) were determined by using GraphPad Prism 8 software.

To evaluate the antiviral efficacy of these drugs, Calu-3 and Vero E6 cells were plated in 48-well cell culture Petri dishes at a density of $5 \times 10^4$ cells/well. Cells were inoculated with IAV (MOI of 1) for 1 h and rocked manually every 10 min to redistribute the inoculum. At 24 h post-infection (hpi), the cells were further treated with the same 10-point dilutions of drugs as described above. Supernatants were collected and processed by heat inactivation (10 min at 95°C) or stored at -80°C until use. Then, viral RNA levels in the supernatant of infected cells were quantified by quantitative reverse transcription PCR (RT-qPCR), representing the quantity of IAV replicated and secreted from cells[44]. A standard curve of 1:5 dilutions of IAV PCR target fragments from $1.5 \times 10^{10}$ to $7.5 \times 10^{5}$ copies/ml was used to quantify the viral RNA with specific primers (fwd, CAAGCAGCAGAGGCCATGGA; rev, GACCAGCACTGGAGCTAGGA). Additionally, the infected cells were harvested with TRIzol, and total cellular RNA was extracted, reverse transcribed into cDNA and subjected to qPCR as previously described[45]. mRNA levels of MX1, ISG15, IFNB1, ACE2 and the housekeeping gene GAPDH were measured with specific primers (*MX1*: fwd, GTTTCCGAAGTGGACATCGCA, rev, CTGCACAGGTTGTTCTCAGC; *ISG15*: fwd, CGCAGATCACCCAGAAGATCG, rev, TTCGTCGCATTTGTCCACCA; *IFNB1*: fwd, GCTTGGATTCCTACAAAGAAGCA, rev, ATAGATGGTCAATGCGGCGTC; *ACE2*: fwd, ACAGTCCACACTTGCCCAAAT, rev, TGAGAGCACTGAAGACCCATT; and *GAPDH*: fwd, GGAGCGAGATCCCTCCAAAAT, rev, GGCTGTTGTCATACTTCTCATGG). Fold changes were calculated by using the ΔΔCT method compared with untreated noninfected cells. Three technical replicates were used for each sample.

## RESULTS

### *Study design and analysis of single-cell data*

Our study highlighted the identification of different therapeutic effects during various disease courses by using publicly available single-cell RNA sequencing data. With the high variability of the cellular compartments underlying disease progression, our drug repurposing profiles from major cell subtypes included T, B, NK, epithelial cells, and macrophages. A total of 9 scRNA-seq BALF samples, including 3 healthy cases, 3 mild cases, and 3 severe cases, were collected from publicly available scRNA-seq data (Supplemental Table S1). After quality filtering, approximately 250,000 gene expression values from 43,914 cells were obtained. The clustering analysis identified six major clusters of macrophages, NK cells, CD4+ T cells, CD8+ T cells, B cells, and epithelial cells (Supplemental Figure S1), which were determined based on the unique signature genes CD68 (macrophage cell), IL7R, CD4 (CD4+ T cell), CD8A (CD8+ T cell), MS4A1 (B cells), and TPPP3 (epithelial cells), respectively (Supplemental Figure S2). We then compared these six major clusters across the healthy, mild, and severe COVID-19 cases and identified differentially expressed genes (DEGs) between any of the two courses, as summarized in Supplemental Table S2. Briefly, the mild-vs-healthy comparison (Supplemental Table S3) had 439 (CD8+ T cell) ~ 1,639 (B cell) DEGs, while the severe-vs-healthy comparison (Supplemental Table S4) had 255 (CD8+ T cell) ~ 1,127 (epithelial cell) and the severe-vs-mild comparison (Supplemental Table S5) had 467 (CD8+ T cell) ~ 1,132 (macrophage cell).

***An overview of drug repurposing via the LINCS database***
Connecting to the LINCS database of small-molecule perturbations of gene expression, we identified candidate drugs and compounds that can reverse the upregulation and downregulation of these genes via the CLUE platform. The closer the connectivity score (CS) is to -100, a score indicating a complete reversal, the higher the chance of identification of drug-adverse effect associations with upregulated or downregulated DEGs. In other words, drugs may show a better response to reverse expression of DEGs upregulated or downregulated in major cell subtypes in the BALF. A total of 281 candidates were selected by CLUE with a CS lower than -90 based on DEGs among all three comparisons between two courses (Supplemental Table S6). These candidates include potential anticoronavirus agents, focusing on FDA-approved drugs and experimental agents that have already been tested in clinical trials. To prioritize known drugs for preclinical and clinical evaluation of the therapeutic effect of SARS-CoV-2, we selected candidates shared by at least three cell types and in ongoing COVID-19 clinical trials, as listed in Tables 1-3.

*Repurposing analysis in mild COVID-19 patients*
To search for candidates for mild cases, we ranked drugs and compounds according to their CSs (Supplemental Table S7). A total of 133 candidate drugs were found to be a potential candidate in at least one cell type among the mild-vs-healthy group, and 53 of them were included in more than one cell subtype (Figure 2A, Supplemental Table S8). The ten drugs identified in three or more cell types included tubulin inhibitors (flubendazole, mebendazole, nocodazole, and vincristine), DNA methyltransferase inhibitor (azacytidine), BCL inhibitor (ABT-737), M5 modulator (VU-0365114-2), calcium channel blocker (calmidazolium), apoptosis stimulant (kinetin-riboside), and opioid receptor antagonist (JTC-801). Six additional drugs are currently undergoing COVID-19 clinical trials (Table 1), including the HIV protease inhibitor lopinavir/ritonavir[46] combination (phase 4), glucocorticoid receptor agonist dexamethasone (phase 3/4)[47], DNA replication inhibitor niclosamide (phase 2/3), antineoplastic agent lenalidomide (phase 4), and calcineurin inhibitor tacrolimus (phase 3)[48].

*Repurposing analysis in severe COVID-19 patients*
 A total of 60 drugs were also found to have the potential to be effective in severe cases compared to in controls (severe vs. healthy group) according to their average CS between the replicates, and 25 of them were involved in more than one cell subtype (Figure 2B, Supplemental Tables S9 & S10). As listed in Table 2, nine drugs presented in at least three separate cell types, including ABT-737 (BCL inhibitor), brefeldin-a (protein synthesis inhibitor), indirubin (CDK inhibitor), TPCA-1 (IKK inhibitor), lopinavir (HIV protease inhibitor), GW-441756 (growth factor receptor inhibitor), treprostinil (prostacyclin analog), tyrphostin-AG-1478 (EGFR inhibitor) and epoxycholesterol (LXR agonist). In this group, lopinavir/ritonavir and hydrocortisone are ongoing in COVID-19 clinical trials.

*Repurposing analysis in severe COVID-19 patients compared to mild patients*
A total of 111 candidate drugs were identified in severe cases compared to mild cases (severe-vs-mild group), 39 of which were involved in more than one cell subtype (Figure 2C, Supplemental Tables S11 & S12). As listed in Table 3, nine drugs (those for which drugs were selected in three separate cell types or more), including fostamatinib (SYK inhibitor), VER-155008 (HSP inhibitor), KU-0063794 (MTOR inhibitor), PIK-90 (PI3K inhibitor), linsitinib (IGF-1 inhibitor), TAK-715

(p38 MAPK inhibitor), Y-27632 (Rho-associated kinase inhibitor), AZ-628 (RAF inhibitor) and lestaurtinib (FLT3 inhibitor), were identified. In this group, except lopinavir, we also assessed the 8 listed drugs in clinical trials for the treatment of COVID-19 in Table 3, including the insulin sensitizer metformin (phase 3), HMGCR inhibitor atorvastatin (phase 2/3), phosphodiesterase inhibitor sildenafil (phase 3), calcium channel blocker verapamil (phase 2/3), phosphodiesterase inhibitor sildenafil (phase 1/2/3), HMGCR inhibitor rosuvastatin (phase 3), CC chemokine receptor antagonist maraviroc (phase 1/2), and dipeptidyl peptidase inhibitor sitagliptin (phase 2/3).

*Shared candidates based on all three comparisons*
As shown in Figures 3A & 3B and Supplemental Table S6, lopinavir was the only one identified in all three comparisons (mild vs healthy, severe vs healthy, and severe vs mild), and interestingly, ritonavir was shared in the mild-vs-healthy and severe-vs-healthy comparisons. There were 23 additional drugs identified in all three comparisons, including SB-216763, ABT-737, JTE-907, brefeldin-a, PKCbeta-inhibitor, indirubin, GW-441756, flubendazole, tyrphostin-AG-1478, memantine, calyculin, kinetin-riboside, ascorbyl-palmitate, ON-01910, mirin, verrucarin-a, emetine, TPCA-1, RHO-kinase-inhibitor-III[rockout], PD-158780 and NVP-AUY922. For example, the glycogen synthase kinase (GSK) inhibitor SB-216763 acts as a neuroprotectant[49] and prevents cardiac ischemia[50]. JTE-907 is a cannabinoid receptor inverse agonist that produces antiinflammatory effects[51].

To further demonstrate the usefulness of the drug repurposing strategy, we identified potential therapeutic drugs based on the transcriptional changes in BALF of COVID-19 patients obtained by using bulk RNA-seq data[40]. Ten candidate drugs were identified using the same analysis pipeline, two of which, including the GSK inhibitor SB-216763 and PPAR receptor antagonist GW-6471, were also included in the single cell-based candidate lists (Supplemental Table S13).

**Adverse drug reaction analysis**

To prioritize the candidates for COVID-19 treatment, we characterized these candidates' known ADRs, which are a central consideration during drug development[52]. We conducted a computational approach using the SEP-L1000 database to predict relationships between drugs and the emergence of ADRs (Supplemental Tables S14 & S15). Figure 4 shows a heatmap of the top 50 drug-ADR associations for on-label (Figure 4A) and off-label (Figure 4B) ADRs. Interestingly, the majority of the candidate drugs were shown with few ADRs. Only lopinavir (associated with 7 off-label ADRs), ritonavir (13 off-label ADRs), and memantine (unexpectedly, 44 on- and 8 off-label ADRs) were included in two or more comparisons. In addition, for mild cases, mebendazole (6 off-label), vincristine (31 on-label), dexamethasone (20 on-label and 37 off-label) and lenalidomide (18 off-label) showed drug-associated ADRs; for severe cases, there were two drugs, treprostinil (35 on- and 6 off-label) and valproic acid (44 on-label); and for comparison between mild and severe cases, metformin was associated with 31 on- and 25 off-label ADRs, atorvastatin 32 off-label, sildenafil 44 on-label, verapamil 32 on- and 27 off-label and dasatinib 45 on- and 22 off-label ADRs. These findings highlighted drug-ADR associations and may lead to informed clinical decisions regarding treatments for COVID-19.

***Potent antiviral activity of candidates against ssRNA viruses in cell culture***
Two candidates, tyrphostin-AG-1478 (AG-1478) and brefeldin-a (BFA), were selected as our top candidates and examined for potent activity against the ssRNA virus influenza A virus (IAV; strain Puerto Rico/8/1934 (H1N1)). IAV is a negative-sense ssRNA virus with eight genomic segments of different lengths (ranging from 0.89 to 2.3 kb). IAV was chosen because both SARS-CoV-2 and IAV are contagious ssRNA viruses that cause respiratory tract infection. AG-1478, an epidermal growth factor receptor (EGFR) inhibitor and a cancer chemotherapy agent, inhibits hepatitis C virus and encephalomyocarditis virus in cells[53]. BFA blocks the envelopment and egress of a mature viral particle by inhibiting protein transfer from the endoplasmic reticulum to the cis-Golgi[54]. BFA treatment inhibits the egress of both DNA viruses (herpes simplex virus)[54] and

RNA viruses (Newcastle disease virus)[54], the entry of human papillomavirus and polyomavirus[55], and the replication of mouse hepatitis coronavirus[56] and human immunodeficiency virus type 1[57].

We tested whether these two drugs reduce viral RNA levels in Calu-3 and Vero E6 cells after infection with IAV by measuring IAV viral RNA levels in the cell culture supernatant by RT-qPCR. Upon treatment, both AG-1478 and BFA reduced IAV viral RNA levels dose dependently at 24 h post-infection (Figure 5A, Supplemental Figure S3A-C). Neither compound caused significant cytotoxicity, but BFA slightly reduced viability in Vero E6 cells at high concentrations. Consistent with the reduction in viral replication, we observed that AG-1478 treatment elevated the mRNA levels of interferon-stimulated genes (ISGs) (ISG15 and MX1) and IFNβ in Calu-3 cells (Figure 5B). The cellular receptor angiotensin-converting enzyme 2 (ACE2) is a key mediator of SARS-CoV-2 host cell entry[58]. We found that cellular mRNA levels of ACE2 increased in AG-1478-treated Calu-3 cells. In contrast, BFA treatment significantly reduced the expression of ACE2, MX1, ISG15, and IFNβ (Supplemental Figure S3D). These results showed that AG-1478 and BFA exhibit a good antiviral effect on IAV, which provided novel possibilities for further preclinical development research and validated clinical therapeutic applications.

**DISCUSSION**

COVID-19 has spread rapidly worldwide; however, no proven vaccine or drug has yet been developed. Various types of drugs, including antivirals, small-molecule drugs, biologics, and vaccines[4, 5], could potentially be used to control or prevent emerging coronavirus disease. However, due to the lack of effective therapeutic agents and long development cycles of vaccines, it is reasonable to consider repurposing existing drugs and compounds for COVID-19 as an alternative approach.

Both IAV and CoVs are common respiratory viruses. The flu season occurs annually, and its symptoms are similar to the respiratory symptoms caused by coronaviruses. In the United States, IAV has caused approximately

9.2 to 35.6 million illnesses with a mortality rate of 0.04%-0.83%[59], while COVID-19 has caused more than 8 million positive cases and 220,000 deaths, with mortality rates of approximately 2.7%[1]. Although COVID-19 has higher morbidity and mortality than IAV infections, plasma cells were increased significantly in both COVID-19 and IAV-infected patients[60]. T cells and NK cells also were activated in both types of patients, which may contribute to defense against the viruses[60]. SARS-CoV-2 needs to be manipulated in BSL-3 conditions for biosafety consideration, while certain IAV strains allow rapid assessment of antiviral activity in vitro in a BSL-2 laboratory.

Our approach is different from previous methods[33-36] for drug repurposing for coronavirus, since it does not merely rapidly identify likely effective therapeutic agents in preventing or treating COVID-19 but tries to filter specific medications targeting immune reactions and specific modulators during the patients' disease courses. Furthermore, the scRNA-seq and transcriptome data were derived from human COVID-19 patients at different disease states (mild vs severe), and two independent publicly available datasets were used for this study.

Several top-scoring drugs out of the 281 drugs we identified have already shown antiviral potential and are even undergoing clinical trials. The tubulin inhibitor flubendazole, widely used in treating intestinal parasites, is a potent inducer of autophagy initiation and can block HIV transfer from dendritic cells to T cells[61]. Azacytidine partially reversed aberrant DNA methylation. Azacytidine combined with chemotherapy (fludarabine and cytarabine) in treating childhood leukemia is in a phase 1 clinical trial, and azacytidine in conjunction with APR-246 for myelodysplastic syndrome is in a phase 3 clinical trial[62]. The BCL inhibitor ABT-737 exhibits potential proapoptotic and antineoplastic activities[63, 64]. The protein synthesis inhibitor brefeldin-a has been widely used to inhibit the entry of some viruses, such as human papillomavirus and polyomavirus[65], and egress of others, such as herpesviruses and paramyxoviruses[66]. Indirubin, an active ingredient of the traditional Chinese medicine "Danggui Longhui Wan", has potent activity against myelocytic leukemia[67] and therapeutic potential as an antiviral agent against IAV[68]. The SYK inhibitor fostamatinib produced clinically meaningful responses for adult persistent and chronic immune thrombocytopenia in two parallel, phase 3 randomized trials[69]. Fostamatinib is now in a phase 2 COVID-19 trial owing to reducing the protein abundance of mucin-1, a biomarker of acute lung injury and respiratory distress syndrome[70]. The HSP inhibitor VER-155008 could regulate Kaposi's sarcoma-associated herpesvirus lytic replication, highlighting its potential as a novel antiviral agent[71]. The FLT3 inhibitor lestaurtinib has secured orphan drug approval from the FDA for acute myeloid leukemia[72] and is in a phase 2 trial for advanced multiple myeloma and phase 1 trials for prostate cancer.

  We also explored the underlying risk factors associated with some side effects of the candidates. Understanding toxic on- and off-label ADRs linked to drugs affecting vital organs can improve patient safety and reduce financial costs[73]. Although the mechanisms of ADRs are complicated and not well understood, SEP-L1000 provides insights into the connection between general structural information and ADRs[74]. Notably, our candidate drugs did not show many ADRs.

  The selection and validation of AG-1478 and BFA were performed because 1) both were shared in the drug repurposing analysis of mild and severe COVID-19 patients; 2) both were expected to be effective in multiple immune cell types (Figure 3B); 3) they can inhibit viruses according to recent reports[53-57]; and 4) no adverse side effects were found in our analyses. These two candidates also showed robust antiviral activities against IAV in vitro in cell lines. Interestingly, AG-1478 or BFA treatment enhanced or inhibited the expression of IFNβ and ISGs, respectively. It is known that type I interferons (IFN-Is) and ISGs confer antiviral activities to host cells. However, inappropriate IFN-Is and ISGs at the wrong time results in excessive inflammation and tissue damage, especially for severe or critical COVID-19[75]. Although ACE2 mediates the cell entry of coronavirus, it also offers protection in acute lung injury against SARS-CoV-2[76]. Recombinant human ACE2 therapy to prevent S-protein interactions with endogenous ACE2 is currently in a phase 2 clinical trial in Europe[77]. Thus, these candidates have therapeutic potential for rational selection to target mild or severe COVID-19 patients. Overall, our data serve as a basis for future pharmacological studies in the treatment of SARS-CoV-2 and will guide the future development of therapies for the different severities of COVID-19 and other viral respiratory infections.

This study may also possess several limitations. First, the public scRNA-seq data had a small number of clinical samples (*n*=9) without detailed patient information available, making comparisons between studies difficult and making it impossible to dissect the patient's clinical features with the particular cell type-mediated immune responses. Second, IAV was chosen to test candidate drug antiviral activity; however, although similar, the immune responses and gene expression patterns during SARS-CoV-2 infection are not the same as those during IAV infection. Future work on large-scale data mining and experimental validation in COVID-19 models would help us better identify antiviral drugs.

## CONCLUSIONS

The COVID-19 pandemic represents the greatest global public health crisis. To date, no proven vaccines or therapies have been developed. We investigated potentially repurposable candidates for the treatment of COVID-19 progression. The findings can guide additional repurposing studies tailored to different stages of disease progression.

## AUTHOR CONTRIBUTIONS

KG, MW, CL, and JH designed the project and revised the manuscript. ZW and KG collected data, performed the analyses, prepared figures, and wrote the manuscript. PG performed analyses, prepared tables, and wrote the manuscript. QP revised the manuscript.

## ACKNOWLEDGMENTS

The work was partially supported by the National Institutes of Health grants P20GM113123 to JH, R01AI138203 and AI109317 to MW and the Science and

## COMPETING INTERESTS

The authors declare that they have no competing interests.

## CODE AND DATA AVAILABILITY

All the codes used in this study are available at https://github.com/guokai8/COVID19/. The scRNA-seq and bulk-RNA-seq data are available from the NCBI Gene Expression Omnibus database (https://www.ncbi.nlm.nih.gov/geo/; accession ID: GSE145926) and the Chinese National Genomics Data Center (https://bigd.big.ac.cn/; accession ID: CRA002390), respectively.

## LIST OF ABBREVIATIONS

ADRs: adverse drug reactions

BALF: bronchoalveolar lavage fluid

CLUE: connectivity map-linked user environment

COVID-19: coronavirus disease 2019

CS: connectivity score

DEGs: differentially expressed genes

FAERS: FDA adverse event report system

HIV: human immunodeficiency virus

ICTRP: international clinical trials registry platform

LINCS: Library of Integrated Network-Based Cellular Signatures

MERS-CoV: Middle East respiratory syndrome coronavirus

PCA: principal component analysis

SARS-CoV: severe acute respiratory syndrome coronavirus

SARS-CoV-2: severe acute respiratory syndrome coronavirus 2

scRNA-seq: single-cell RNA sequencing

SEP-L1000: side effect prediction based on L1000

ssRNA: single-stranded ribonucleic acid

IFN-Is: type I interferons

UMAP: uniform manifold approximation and projection

**FIGURE LEGENDS**

**Figure 1. Workflow of drug repurposing for treating different durations of COVID-19.** Publicly available scRNA-seq data and transcriptomic data of BALF from COVID-19 patients were input against the LINCS database using the CLUE platform. Candidates are selected that can reverse the expression of upregulated DEGs upon drug treatment and were compared by connectivity score and the number of major cell subtypes across healthy, mild, and severe groups. This figure was created by modifying from Servier Medical Art (SMART) and Vecteezy.com.

**Figure 2. UpSet plots showing the overlap among the drug candidates for treating COVID-19 based on the LINCS database**. DEGs between (A) mild and healthy, (B) severe and healthy, and (C) severe and mild samples in B cells, $CD4^+$ T cells, $CD8^+$ T cells, epithelial cells, NK cells, and macrophages.

**Figure 3. Common drug candidates**. Venn diagram showing the overlap among the drug candidates for treating COVID-19 between three sets across the control, mild, and severe COVID-19 groups (A) and heatmap showing the 25 drugs shared by at least two sets (B). MvsH: mild-vs-healthy; SvsH: severe-vs-healthy; SvsM: severe-vs-mile.

**Figure 4. Heatmap of drug-ADR association**. On-label (A) and off-label (B) ADRs are illustrated in heatmaps. White color indicates no association between drug and ADRs.

**Figure 5. Tyrphostin-AG-1478 is effective against IAV in Calu-3 cell cultures**. (A) Antiviral activity of tyrphostin-AG-1478 (AG-1478, 0 to 10 µM, IC50 = 0.043 µM) was assessed in Calu-3 cells infected with IAV PR8 (H1N1, MOI of 1) at 24 hpi. Cell viability is depicted in black. $IC_{50}$ was calculated based on normalization to the control and fitted in GraphPad Prism. (B) Bar plots showing mRNA levels of cellular *ACE2*, *MX1*, *ISG15*, and *IFNB1,* calculated with ΔΔCT

over noninfected Calu-3 cells. Calu-3 cells were treated with AG-1478 (0 to 10 µM) and infected with IAV (MOI of 1). Data are represented as the mean ± SD with 3 technical replicates each. Statistical significance was determined using one-way ANOVA. *$P < 0.05$, **$P < 0.01$, ***$P < 0.001$, ****$P < 0.0001$.

# TABLES

**Table 1.** A list of potential drugs for treating COVID-19 based on the LINCS database and DEGs between mild and healthy samples in B cells, CD4⁺ T cells, CD8⁺ T cells, epithelial cells, NK cells, and macrophages. Connectivity scores were calculated from the CLUE platform. Asterisk (*) represents a clinical trial for its efficacy in COVID-19 disease. (+) indicates drugs meeting the SC < -90 criteria, while (-) indicates drugs not meeting the criterion.

| Drug | LINCS ID | B cells | CD4+ T cells | CD8+ T cells | Epithelial cells | Macrophages | NK cells | Shared Sets | Description | Phase* |
|---|---|---|---|---|---|---|---|---|---|---|
| Flubendazole | BRD-K86003836 | + | + | + | + | - | + | 5 | Tubulin inhibitor | |
| Azacitidine | BRD-K03406345 | - | + | + | + | + | - | 4 | DNA methyltransferase inhibitor | |
| ABT-737 | BRD-K56301217 | + | + | - | - | - | + | 3 | BCL inhibitor | |
| VU-0365114-2 | BRD-K37456065 | - | + | + | + | - | - | 3 | M5 modulator | |
| Calmidazolium | BRD-A98283014 | - | + | - | + | + | - | 3 | Calcium channel blocker | |
| Mebendazole | BRD-K77987382 | - | + | - | + | + | - | 3 | Tubulin inhibitor | |
| Kinetin-riboside | BRD-K94325918 | - | + | - | + | + | - | 3 | Apoptosis stimulant | |
| Nocodazole | BRD-K12539581 | - | + | - | + | + | - | 3 | Tubulin inhibitor | |
| JTC-801 | BRD-K17705806 | - | + | - | + | + | - | 3 | Opioid receptor antagonist | |
| Vincristine | BRD-K82109576 | - | - | + | + | - | + | 3 | Tubulin inhibitor | |
| Lopinavir | BRD-K99451608 | + | - | - | - | - | + | 2 | HIV protease inhibitor | Phase 4 |
| Ritonavir | BRD-K51485625 | + | - | - | - | - | + | 2 | HIV protease inhibitor | Phase 4 |
| Dexamethasone | BRD-A35108200 | - | + | - | - | - | - | 1 | Glucocorticoid receptor agonist | Phase 3/4 |
| Niclosamide | BRD-K35960502 | - | - | - | - | + | - | 1 | DNA replication inhibitor | Phase 2/3 |
| Lenalidomide | BRD-K05926469 | - | - | - | - | - | + | 1 | Antineoplastic | Phase 4 |
| Tacrolimus | BRD-K69608737 | - | - | + | - | - | - | 1 | Calcineurin inhibitor | Phase 3 |

**Table 2.** A list of potential drugs for treating COVID-19 based on the LINCS database and DEGs between severe and healthy samples in B cells, CD4+ T cells, CD8+ T cells, epithelial cells, NK cells, and macrophages. Connectivity scores were calculated from the CLUE platform. Asterisk (*) represents a clinical trial for its efficacy in COVID-19 disease. (+) indicates drugs meeting the SC < -90 criteria, while (-) indicates drugs not meeting the criterion.

| Drug | LINCS ID | B cells | CD4+ T cells | CD8+ T cells | Epithelial cells | Macrophages | NK cells | Shared Sets | Description | Phase* |
|---|---|---|---|---|---|---|---|---|---|---|
| ABT-737 | BRD-K56301217 | + | + | + | - | + | + | 5 | BCL inhibitor | |
| Brefeldin-a | BRD-A17065207 | + | + | + | - | + | + | 5 | Protein synthesis inhibitor | |
| Indirubin | BRD-K53959060 | + | + | + | - | + | + | 5 | CDK inhibitor | |
| TPCA-1 | BRD-K51575138 | + | + | + | - | - | + | 4 | IKK inhibitor | |
| Lopinavir | BRD-K99451608 | + | - | - | - | + | + | 3 | HIV protease inhibitor | Phase 4 |
| GW-441756 | BRD-K04146668 | + | - | - | - | + | + | 3 | Growth factor receptor inhibitor | |
| Treprostinil | BRD-A67438293 | + | - | + | - | - | + | 3 | Prostacyclin analog | |
| Tyrphostin-AG-1478 | BRD-K68336408 | - | + | + | - | + | - | 3 | EGFR inhibitor | |
| Epoxycholesterol | BRD-K61480498 | - | + | + | + | - | - | 3 | LXR agonist | |
| Ritonavir | BRD-K51485625 | - | - | - | - | + | + | 2 | HIV protease inhibitor | Phase 4 |
| Hydrocortisone | BRD-A07000685 | - | - | + | - | - | - | 1 | Glucocorticoid receptor agonist | Phase 3 |

**Table 3.** A list of potential drugs for treating COVID-19 based on the LINCS database and DEGs between severe and mild samples in B cells, CD4+ T cells, CD8+ T cells, epithelial cells, NK cells, and macrophages. Connectivity scores were calculated from the CLUE platform. Asterisk (*) represents a clinical trial for its efficacy in COVID-19 disease. (+) indicates drugs meeting the SC < -90 criteria, while (-) indicates drugs not meeting the criterion.

| Drug | LINCS ID | B cells | CD4+ T cells | CD8+ T cells | Epithelial cells | Macrophages | NK cells | Shared Sets | Description | Phase* |
|---|---|---|---|---|---|---|---|---|---|---|
| Fostamatinib | BRD-K20285085 | + | + | + | + | + | + | 6 | SYK inhibitor | Phase 2 |
| VER-155008 | BRD-K32330832 | + | + | + | + | + | + | 6 | HSP inhibitor | |
| KU-0063794 | BRD-K67566344 | + | - | - | - | + | + | 3 | MTOR inhibitor | |
| PIK-90 | BRD-K99818283 | + | + | - | - | - | + | 3 | PI3K inhibitor | |
| Linsitinib | BRD-K08589866 | + | - | - | - | + | + | 3 | IGF-1 inhibitor | |
| TAK-715 | BRD-K52751261 | - | + | + | - | - | + | 3 | p38 MAPK inhibitor | |
| Y-27632 | BRD-K44084986 | - | + | + | - | - | + | 3 | Rho associated kinase inhibitor | |
| AZ-628 | BRD-K05804044 | - | - | + | + | - | + | 3 | RAF inhibitor | |
| Lestaurtinib | BRD-K23192422 | - | - | - | + | + | + | 3 | FLT3 inhibitor | |
| Metformin | BRD-K79602928 | - | - | + | - | - | - | 1 | Insulin sensitizer | Phase 3 |
| Atorvastatin | BRD-U88459701 | - | - | - | + | - | - | 1 | HMGCR inhibitor | Phase 2/3 |
| Sildenafil | BRD-K50128260 | - | - | - | + | - | - | 1 | Phosphodiesterase inhibitor | Phase 3 |
| Verapamil | BRD-A09533288 | - | - | - | + | - | - | 1 | Calcium channel blocker | Phase 2/3 |
| Lopinavir | BRD-K99451608 | - | - | - | + | - | - | 1 | HIV protease inhibitor | Phase 4 |
| Sildenafil | BRD-K79759585 | - | - | - | + | - | - | 1 | Phosphodiesterase inhibitor | Phase 1/2/3 |
| Rosuvastatin | BRD-K82941592 | - | - | - | + | - | - | 1 | HMGCR inhibitor | Phase 3 |
| Maraviroc | BRD-A04352665 | - | - | - | + | - | - | 1 | CC chemokine receptor antagonist | Phase 1/2 |
| Sitagliptin | BRD-K19416115 | - | - | - | - | + | - | 1 | Dipeptidyl peptidase inhibitor | Phase 2/3 |

**SUPPLEMENTARY INFORMATION**

**Supplementary Figure S1.** UMAP presentation of a single-cell atlas of BALFs showing 6 major cell types.

**Supplementary Figure S2.** The violin plot shows the expression of signature genes (CD68, IL7R, CD4, CD8A, MS4A1, and TPPP3) from major cell types.

**Supplementary Figure S3.** The antiviral activities of tyrphostin-AG-1478 and brefeldin-a against IAV in vitro. (A-C) Antiviral activity and dose-response curves of tyrphostin-AG-1478 (AG-1478, 0 to 10 μM) and brefeldin-a (BFA, 0 to 10 μM) in Calu-3 cells (B) and Vero-E6 cells (A, C) infected with IAV PR8 (H1N1, MOI of 1) at 24 hpi. Cell viability is depicted in black. The $IC_{50}$ was calculated based on normalization to the control and fitted in GraphPad Prism. (D) Bar plots showing mRNA levels of cellular *ACE2*, *MX1*, *ISG15*, and *IFNB1,* calculated with ΔΔCT over noninfected Calu-3 cells. Calu-3 cells were treated with BFA (0 to 10 μM) and infected with IAV (MOI of 1). Data are represented as the mean ± SD with 3 technical replicates each. Statistical significance was determined using one-way ANOVA. **$P < 0.01$, ****$P < 0.0001$.

**Supplemental Table S1.** Clinical data of the enrolled subjects (SARS-COV-2 confirmed).

**Supplemental Table S2.** The numbers of differentially expressed genes in different cell types (clusters) among three group comparisons.

**Supplemental Table S3.** Differentially expressed genes between mild and healthy samples in B cells, CD4+ T cells, CD8+ T cells, epithelial cells, NK cells and macrophages.

**Supplemental Table S4.** Differentially expressed genes between severe and healthy samples in B cells, CD4+ T cells, CD8+ T cells, epithelial cells, NK cells and macrophages.

**Supplemental Table S5.** Differentially expressed genes between severe and mild samples in B cells, CD4+ T cells, CD8+ T cells, epithelial cells, NK cells and macrophages.

**Supplemental Table S6.** Full list of overlapping potential drugs among three comparisons (mild vs healthy, severe vs healthy and severe vs mild).

**Supplemental Table S7.** Full list of potential drugs for treating COVID-19 based on the LINCS database and DEGs between mild and healthy samples in B cells, CD4$^+$ T cells, CD8$^+$ T cells, epithelial cells, NK cells and macrophages. Connectivity scores were calculated from the CLUE platform.

**Supplemental Table S8.** Overlapping potential drugs based on DEGs among different cell clusters between mild and healthy patients.

**Supplemental Table S9.** Full list of potential drugs for treating COVID-19 based on the LINCS database and DEGs between severe and healthy samples in B cells, CD4$^+$ T cells, CD8$^+$ T cells, epithelial cells, NK cells and macrophages. Connectivity scores were calculated from the CLUE platform.

**Supplemental Table S10.** Overlapping potential drugs based on DEGs among different cell clusters between severe and healthy patients.

**Supplemental Table S11.** Full list of potential drugs for treating COVID-19 based on the LINCS database and DEGs between severe and mild samples in B cells, CD4$^+$ T cells, CD8$^+$ T cells, epithelial cells, NK cells and macrophages. Connectivity scores were calculated from the CLUE platform.

**Supplemental Table S12.** Overlapping potential drugs based on DEGs among different cell clusters between severe and mild patients.

**Supplemental Table S13.** A list of potential drugs for treating COVID-19 based on DEGs from RNA-seq data between patient and healthy samples and the LINCS database.

**Supplemental Table S14.** A complete list of on-label ADRs of all candidates. On-label ADRs of drugs were downloaded from the Side Effect Resource (SIDER). The ADR terms are mapped to Preferred Terms (PTs) coded in MedDRA v16.0.

**Supplemental Table S15.** A complete list of off-label ADRs of all candidates. Off-label ADRs were processed data from the postmarket ADR reports within the FDA Adverse Event Report System (FAERS). The ADR terms are mapped to Preferred Terms (PTs) coded in MedDRA v16.0.

# FIGURES

## Figure 1

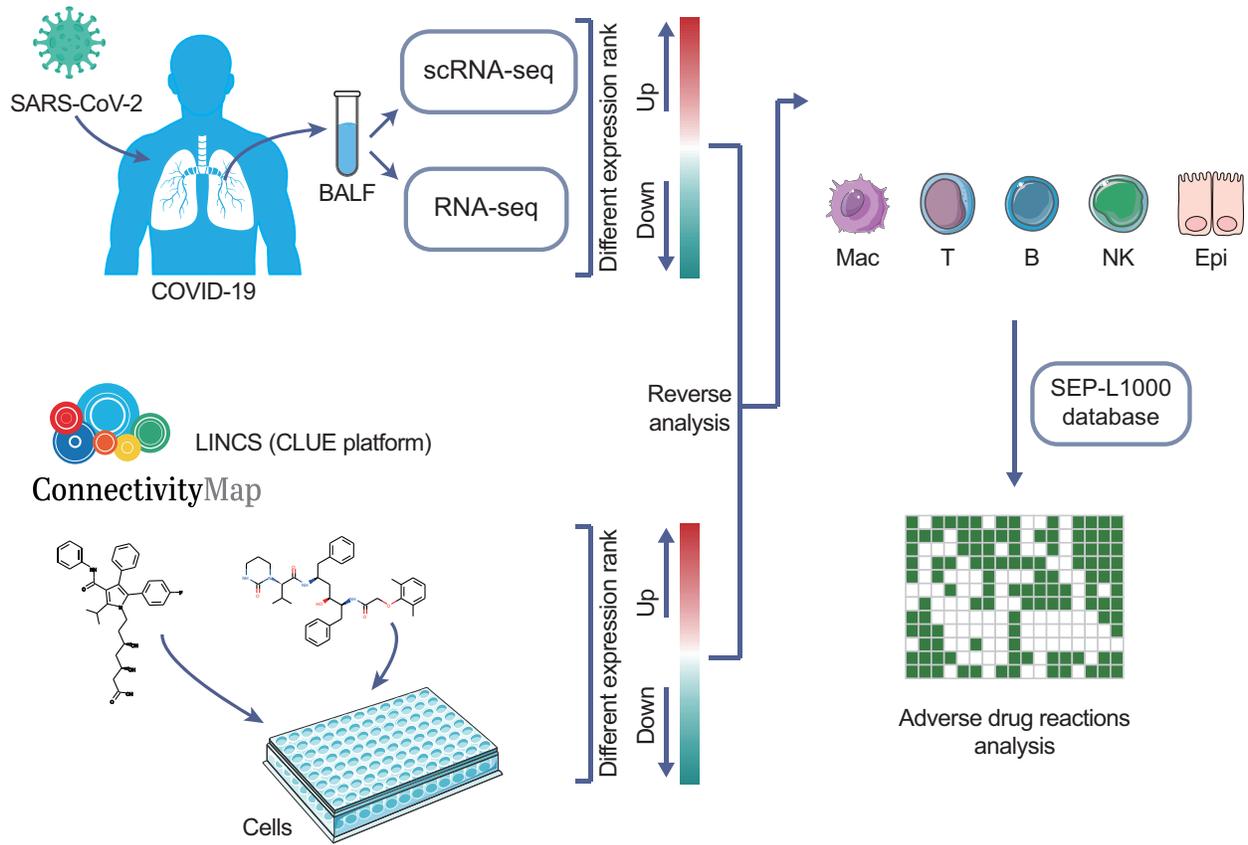

**Figure 2**

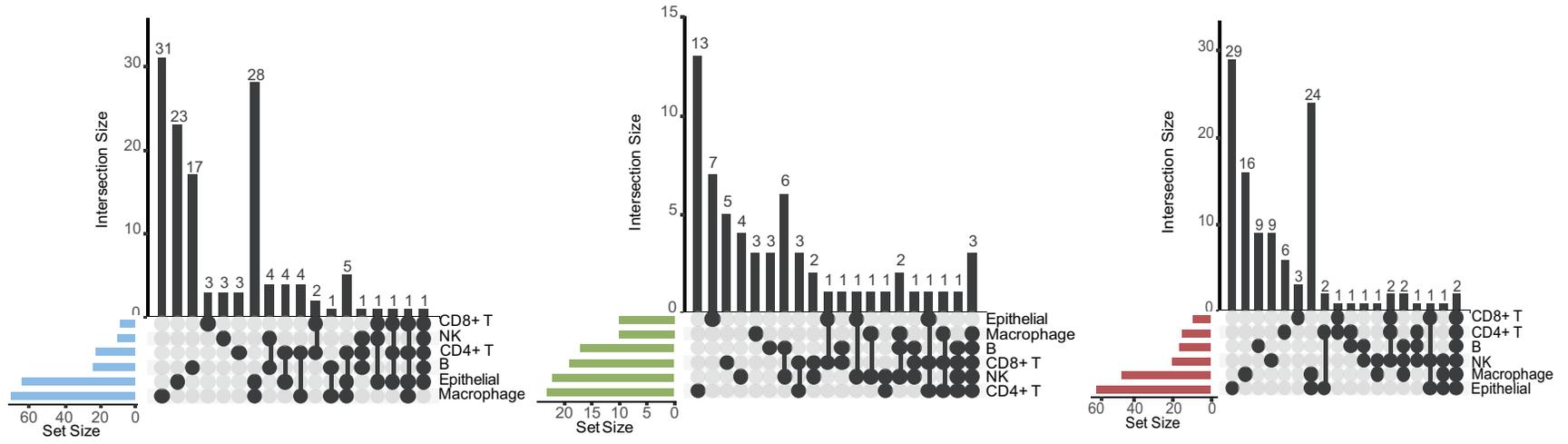

**Figure 3**

**A**

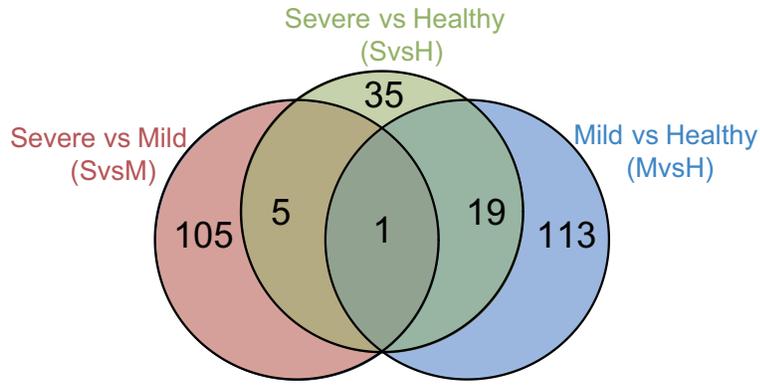

**B**

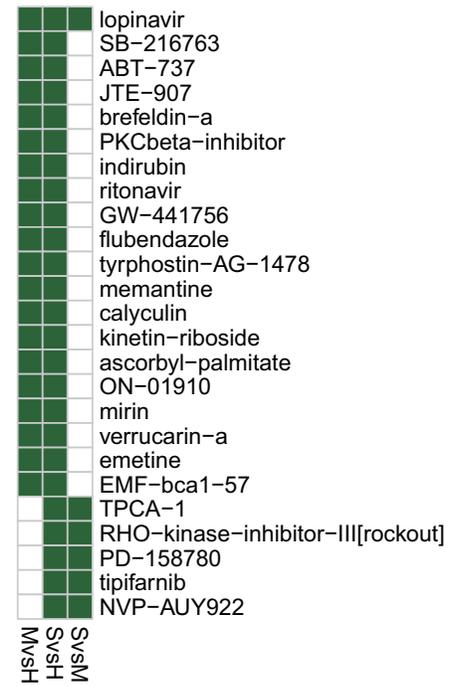

A 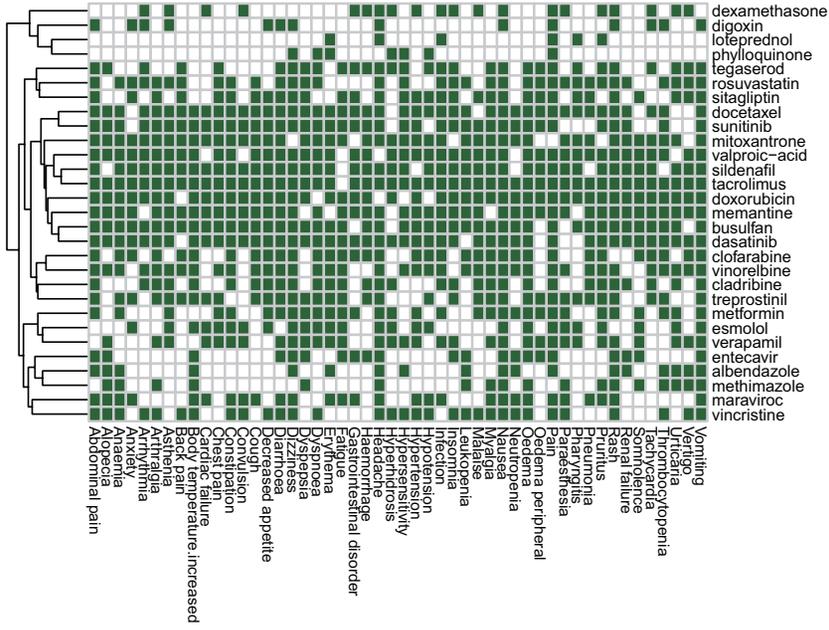

B 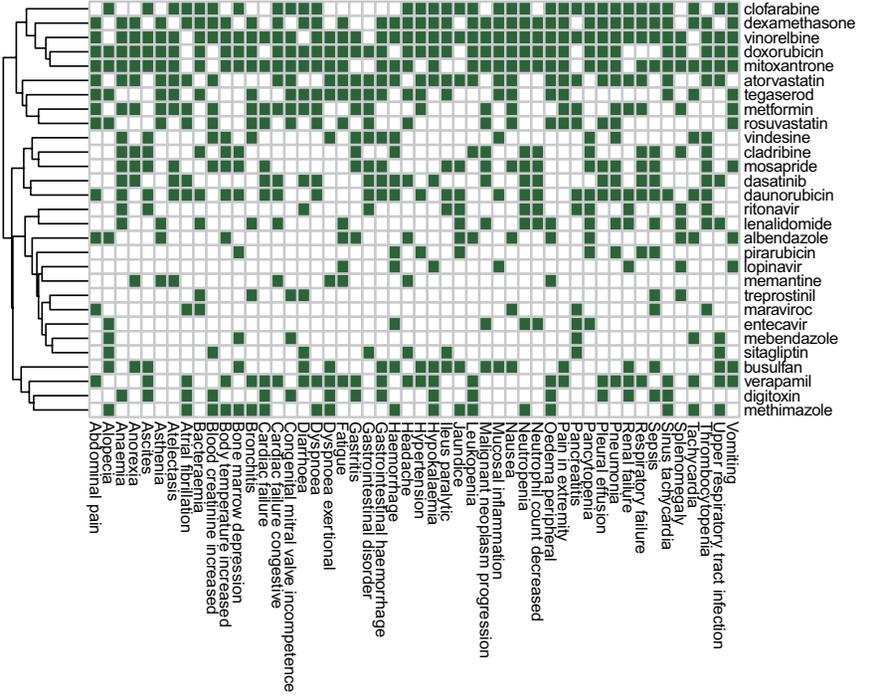

**Figure 5**

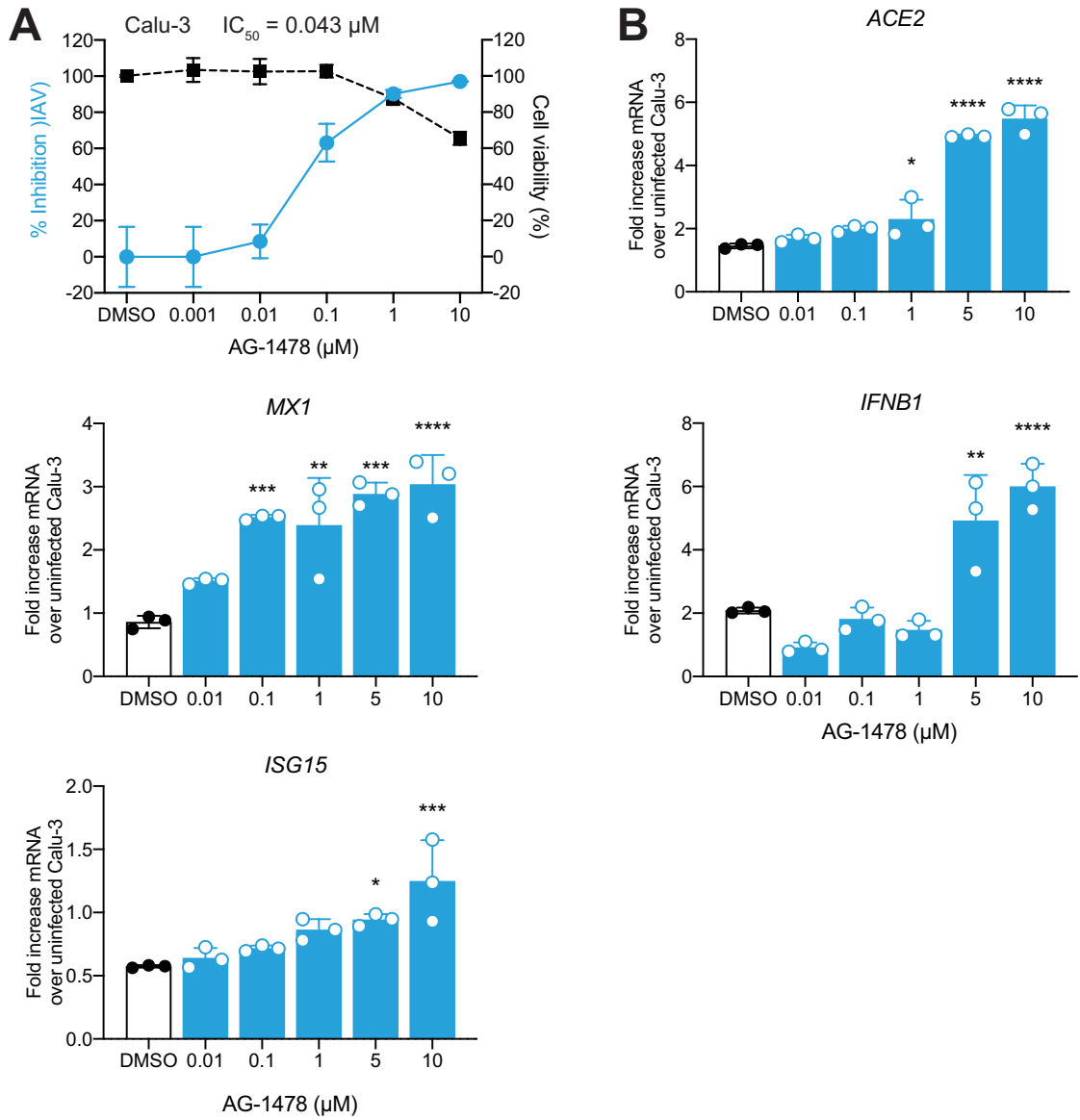

**Supplementary Figure S1**

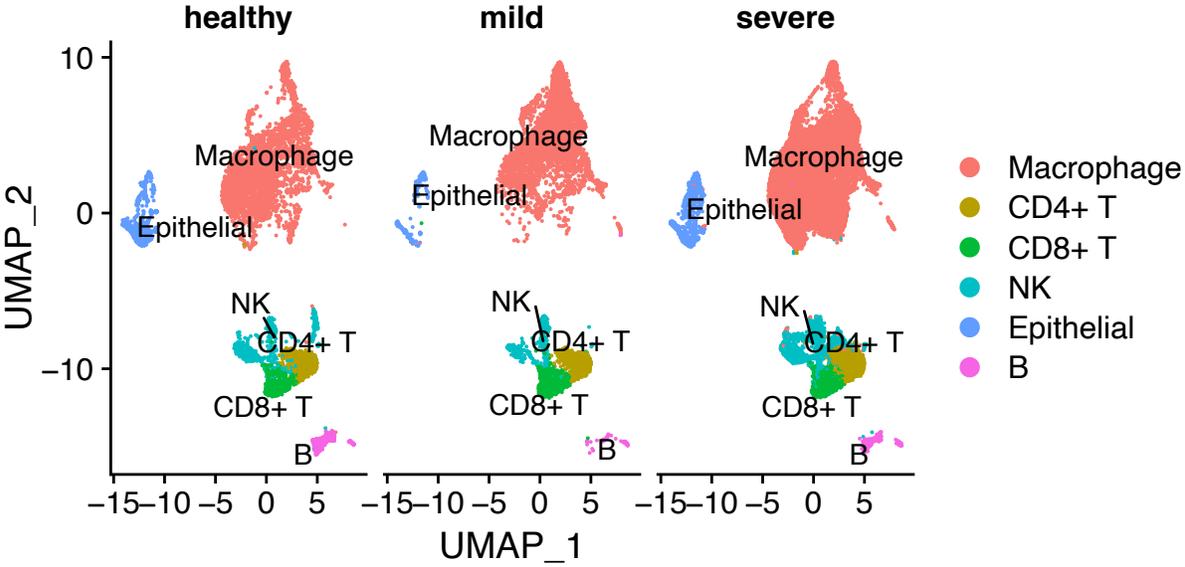

**Supplementary Figure S2**

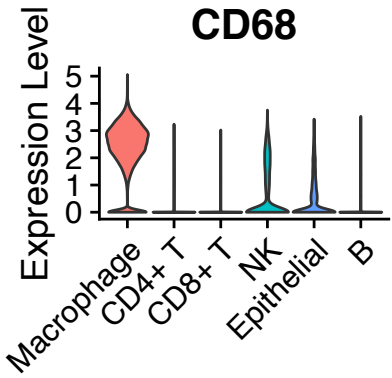 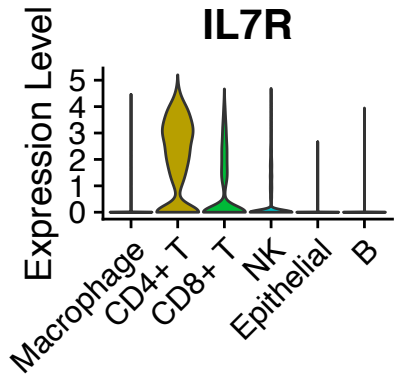 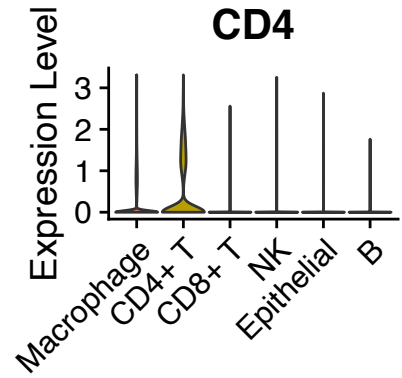
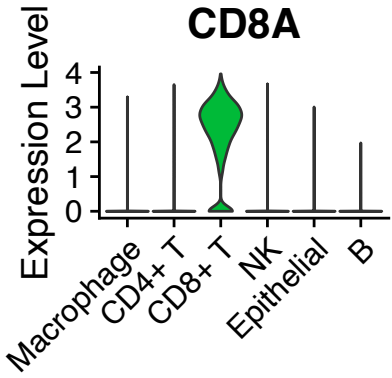 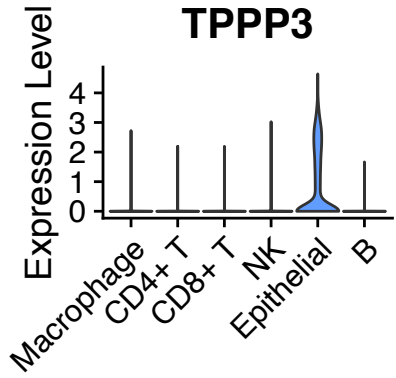 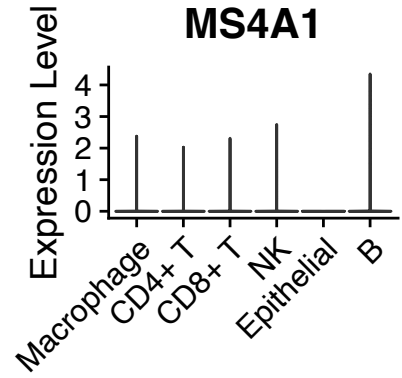

# Supplementary Figure S3

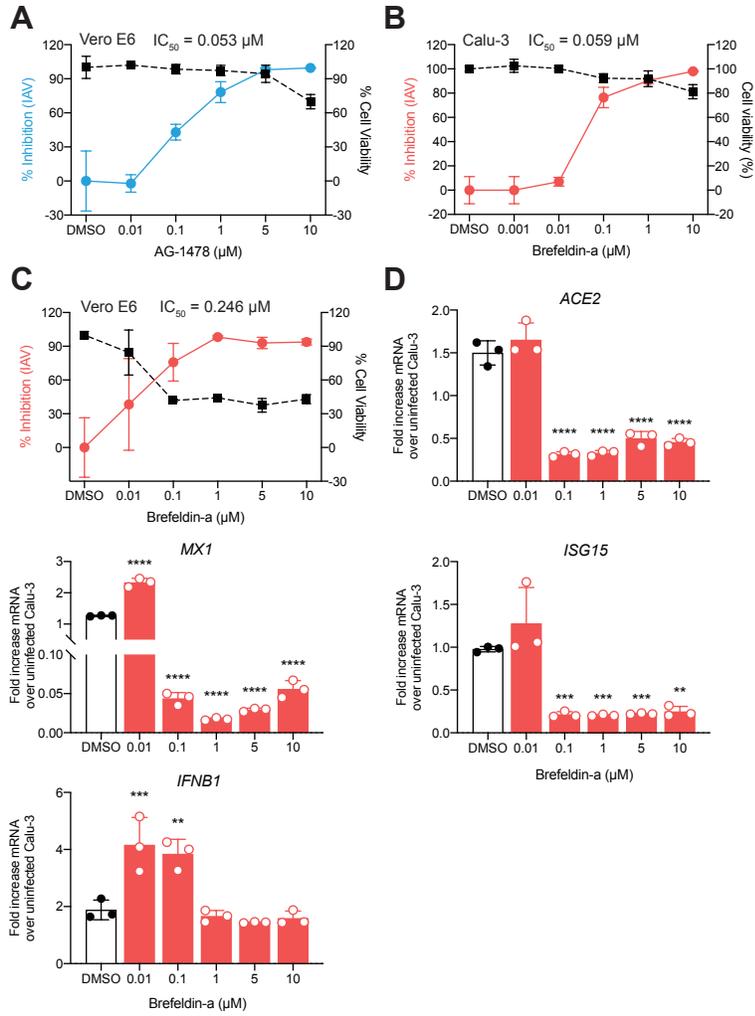